\pdfoutput=1
\documentclass[aps,prd,10pt,onecolumn,nofootinbib,superscriptaddress,floatfix]{revtex4-2}

\usepackage{amsmath,amssymb,graphicx,bm}
\usepackage[hidelinks]{hyperref}

\newcommand{\KernX}{{\cal K}_X}

\newcommand{\GammaS}{\Gamma_S}
\newcommand{\Pieff}{\Pi_{\rm eff}}
\newcommand{\asph}{a_{\rm sph}}
\newcommand{\TR}{T_R}
\newcommand{\TF}{T_F}
\newcommand{\Tov}{T_{\rm ov}}

\begin{document}

\title{Baryogenesis from the Thermodynamic Arrow of Time: a Transfer-Function Bound and an Entropy-Clock Mechanism}

\author{Yakov Mandel}
\email{yakovm2000@proton.me}
\thanks{ORCID: 0009-0006-1766-6695}
\affiliation{Independent Researcher, Haifa, Israel}

\begin{abstract}
We formulate a transfer test for baryogenesis driven by time-dependent derivative sources that generate an effective charge bias.
For a zero-mean oscillatory chemical potential convolved with a smooth finite-time kernel, the frozen asymmetry is low-pass filtered.
In a one-sided exponential effective-kernel benchmark the signed response is
$I_\varphi(x)=(\cos\varphi-x\sin\varphi)/(1+x^2)$, while the phase-optimized envelope is
$F_{\rm amp}(x)=1/\sqrt{1+x^2}$, with $x=\omega\tau_{\rm off}$.
The benchmark's sharp onset produces its $1/x$ high-frequency envelope; smoother turn-ons can yield stronger suppression.
A more general integration-by-parts bound shows that a smooth source with many sign changes across the production-survival kernel is controlled by its residual low-frequency component.
We then study an entropy-clock ansatz, $\theta_X=\epsilon_X\ln(S/S_0)$, which gives $\mu_X=\epsilon_X d\ln S/dt$ during entropy-producing reheating.
The yield equation is written with the required entropy-dilution term.
For a perturbative matter-dominated reheating stage, $S\propto a^{15/8}$ and hence $\Pi=d\ln S/d\ln a=15/8$ before reheating completes.
The crucial consistency condition is that entropy production and charge-violating freeze-out must overlap: in a Weinberg-operator $B-L$ benchmark this requires a near-completion regime $\TR={\cal O}(\TF)$, with $\TF\sim10^{12}$--$10^{13}\,\mathrm{GeV}(0.05\,\mathrm{eV}/\bar m_\nu)^2$.
If $\TR\gg\TF$, entropy production ends before freeze-out and the source is washed out; if $\TR\ll\TF$, the interaction is never efficient during reheating.
Thus the mechanism is not a generic reheating effect, but a conditional coincidence between the reheating scale and the charge-violation scale; its precise width requires a full kinetic solution.
For a direct baryon source the radiation-dominated normalization gives $|\epsilon_X\Pieff|\simeq3.2\times10^{-3}(10^{12}\,\mathrm{GeV}/\Tov)$; for a $B-L$ source reprocessed by sphalerons the required magnitude is larger by $1/\asph=79/28$.
With $\Pi=15/8$ this corresponds to $|\epsilon_X|\zeta\simeq1.7\times10^{-3}$ and $4.8\times10^{-3}$, respectively, at $\Tov=10^{12}\,\mathrm{GeV}$, where $\zeta\le1$ is the overlap efficiency.
The entropy-clock source is treated phenomenologically; a UV completion must explain why the relevant charge-biasing phase or background variable tracks $\ln S$ or an equivalent monotonic dissipative variable.
\end{abstract}

\maketitle

\section{Introduction}
The origin of the baryon asymmetry remains a central problem in particle cosmology.
Sakharov's conditions~\cite{Sakharov:1967dj} require baryon-number violation, C and CP violation, and departure from equilibrium.
Spontaneous baryogenesis~\cite{Cohen:1987vi,Cohen:1988kt} realizes the departure from equilibrium through a time-dependent phase, producing an effective chemical potential $\mu_X=\dot\theta_X$ for a charge $X$.
The cosmological consistency of such scenarios depends not only on the local value of $\mu_X$, but on the expansion history, the charge-violating rates, backreaction, isocurvature constraints, and washout~\cite{DeSimone:2016ofp,Buchmuller:2004nz,Davidson:2008bu}.
Closely related gravitational-baryogenesis constructions face additional dynamical constraints when the baryon current is coupled to curvature derivatives~\cite{Davoudiasl:2004gf,Arbuzova:2017dwn}.

This paper isolates one limited question.
If $\mu_X(t)$ is finite in duration or changes sign during freeze-out, how much of it survives the production-survival kernel?
For a zero-mean oscillatory source the answer is a transfer-function suppression: only the low-frequency component across the active kernel contributes efficiently.
This is not a replacement for a Boltzmann calculation, but a diagnostic that should be applied before interpreting a large microscopic source amplitude as successful baryogenesis.

Nonstationary spontaneous-baryogenesis kinetics, including finite integration intervals, were studied in Ref.~\cite{Arbuzova:2016qfh}.
Our narrower contribution is to cast the source--freeze-out mismatch as a normalized frequency response, include entropy dilution in the production-survival kernel, and apply the same kernel logic to reheating overlap.

Recent work emphasizes finite-duration phase-transition dynamics, bubble collisions, reheating-era generation, and defect evolution~\cite{Ghoshal:2026bubble,An:2026bubble,Dorsch:2026wallgo,Klose:2026resummation,Barman:2026subew,Behera:2026modular,Barbini:2026walls,Cho:2026thdm}.
Oscillating asymmetric-inflaton backgrounds provide another standard setting for a time-dependent charge bias~\cite{Takahashi:2015ula}.
In all such cases, the final asymmetry is controlled by a temporal overlap of source, violation window, washout, and possible entropy dilution.

We then examine a possible way to avoid zero-mean suppression: a sign-definite source proportional to the irreversible entropy-production rate,
\begin{equation}
\theta_X(t)=\epsilon_X\ln\!\left[\frac{S(t)}{S_0}\right],
\qquad
\mu_X(t)=\dot\theta_X=\epsilon_X\frac{d\ln S}{dt},
\qquad S=a^3s .
\label{eq:entropy_clock_intro}
\end{equation}
This entropy-clock form is an ansatz, not a theorem.
The second law supplies monotonicity of $S$ during entropy production, but microphysics must supply a phase or background with the appropriate C and CP transformation properties that tracks $\ln S$ or a related monotonic dissipative variable.
The nontrivial point is that the same epoch must also contain active charge violation.
For the Weinberg-operator benchmark, this becomes the order-unity overlap requirement $\TR={\cal O}(\TF)$.
A related gravitational extension of the entropy-clock idea was explored in Ref.~\cite{Mandel:2026unified}; the present analysis isolates the kinetic consistency test and treats the entropy-clock source phenomenologically.

\section{Rate equation, dilution, and transfer bound}
Let $X$ be the charge whose asymmetry is produced, for example $B$ or $B-L$.
A derivative coupling $(\partial_\mu\theta_X)J_X^\mu$ gives a chemical potential $\mu_X=\dot\theta_X$. A homogeneous background with $\dot\theta_X\neq0$ selects the cosmological plasma frame and acts as a spontaneous CPT-breaking charge bias~\cite{Cohen:1987vi,Cohen:1988kt}. The CP transformation properties of $\theta_X$, and therefore the microscopic realization of Sakharov CP violation, are model dependent and must be specified by a concrete completion.
For $|\mu_X|/T\ll1$,
\begin{equation}
Y_X^{\rm eq}\equiv \frac{n_X^{\rm eq}}{s}=\kappa_X\frac{\mu_X}{T},
\qquad
\kappa_X\equiv \frac{15\chi_X}{4\pi^2g_*} .
\label{eq:kappa_def}
\end{equation}
The numerical value $\kappa_X=0.02$ used below corresponds to $\chi_X\simeq5.3$ for $g_*=100$.
When comoving entropy changes,
\begin{equation}
\GammaS(t)\equiv \frac{d\ln S}{dt},
\end{equation}
the yield obeys
\begin{equation}
\dot Y_X+\bigl[\Gamma_X(t)+\GammaS(t)\bigr]Y_X
=\Gamma_X(t)\kappa_X\frac{\mu_X(t)}{T(t)} .
\label{eq:yield_with_entropy}
\end{equation}
The late-time solution is
\begin{equation}
Y_X(\infty)=\kappa_X\int dt\,\KernX(t)\frac{\mu_X(t)}{T(t)},
\label{eq:master_kernel}
\end{equation}
with
\begin{equation}
\KernX(t)=\Gamma_X(t)\exp\!\left[-\int_t^\infty
\bigl(\Gamma_X(t')+\GammaS(t')\bigr)dt'\right].
\label{eq:kernel_def}
\end{equation}
Thus $\KernX$ contains both washout and entropy dilution.
For an oscillatory source
\begin{equation}
\frac{\mu_X(t)}{T(t)}=q_0A(t)\cos(\omega t+\varphi),
\end{equation}
Eq.~\eqref{eq:master_kernel} becomes
\begin{equation}
Y_X(\infty)=\kappa_Xq_0\,\mathrm{Re}\left[e^{i\varphi}
\int dt\,\KernX(t)A(t)e^{i\omega t}\right].
\label{eq:fourier_kernel}
\end{equation}
If $f(t)=\KernX(t)A(t)$ is smooth, integrable, and vanishes at the endpoints, integration by parts gives
\begin{equation}
\left|Y_X(\infty)\right|
\le
\frac{\kappa_X|q_0|}{\omega}
\int dt\,\left|\frac{d}{dt}f(t)\right| .
\label{eq:ibp_bound}
\end{equation}
This is the general transfer statement: a source with many sign changes across a smooth kernel is suppressed unless the kernel has sharp features at the same time scale or the source contains a nonzero DC component.

For the solvable exponential benchmark, define the normalized effective kernel-envelope product
\begin{equation}
\widehat f_{\tau}(t)\equiv
\frac{\KernX(t)A(t)}{\int_0^\infty dt'\,\KernX(t')A(t')}
=\frac{1}{\tau_{\rm off}}e^{-t/\tau_{\rm off}}\Theta(t).
\label{eq:effective_exponential}
\end{equation}
Here $\tau_{\rm off}$ is the effective width of the combined production-survival kernel and source envelope. The normalized signed response is then
\begin{equation}
I_\varphi(x)=\int_0^\infty dt\,\widehat f_{\tau}(t)\cos(\omega t+\varphi)
=\frac{\cos\varphi-x\sin\varphi}{1+x^2},
\qquad
x\equiv\omega\tau_{\rm off}.
\label{eq:phase_transfer}
\end{equation}
The phase-optimized envelope is
\begin{equation}
F_{\rm amp}(x)=\max_\varphi |I_\varphi(x)|=\frac{1}{\sqrt{1+x^2}} .
\label{eq:Fx}
\end{equation}
The signed result for a fixed phase can be smaller than this envelope or have the opposite sign.
Because the one-sided exponential benchmark has a sharp onset at $t=0$, its $1/x$ high-frequency envelope is boundary dominated.
A smoother turn-on can produce stronger asymptotic suppression.

\begin{figure}[t]
\centering
\includegraphics[width=0.88\columnwidth]{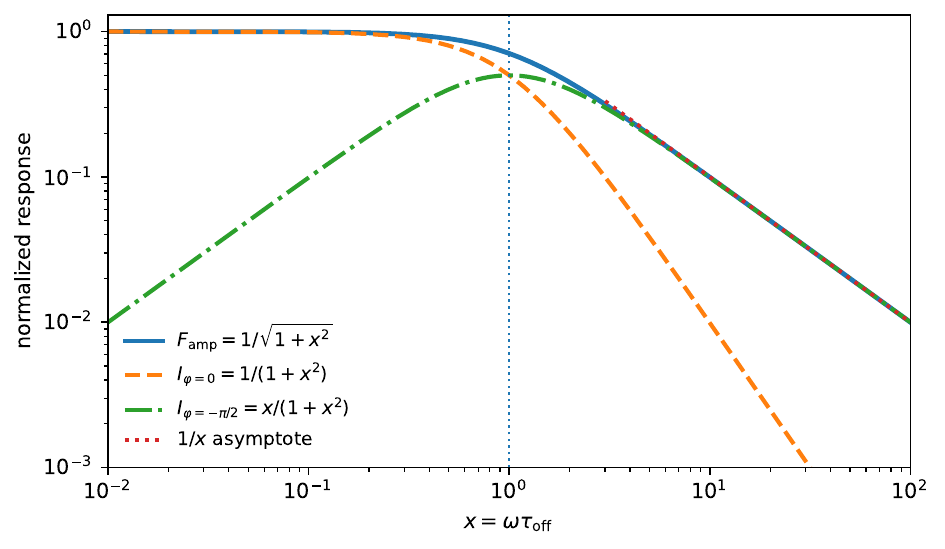}
\caption{Transfer response for the one-sided exponential effective-kernel benchmark. The solid curve is the phase-maximized envelope $F_{\rm amp}(x)=(1+x^2)^{-1/2}$. Signed responses are given by Eq.~\eqref{eq:phase_transfer}. The dotted $1/x$ asymptote is drawn only for $x\gtrsim3$, where the sharp-onset asymptotic comparison is meaningful; smoother turn-ons may fall faster.}
\label{fig:F}
\end{figure}

\section{Entropy-clock source and the reheating overlap}
The entropy-clock source can be written as
\begin{equation}
\mu_X=\epsilon_X\frac{d\ln S}{dt}=\epsilon_XH\Pi,
\qquad
\Pi\equiv\frac{d\ln S}{d\ln a} .
\label{eq:entropy_clock}
\end{equation}
In perturbative matter-dominated reheating with a subdominant radiation bath, $T\propto a^{-3/8}$ before completion and hence $S=a^3s\propto a^3T^3\propto a^{15/8}$~\cite{Allahverdi:2010xz}.
Therefore
\begin{equation}
\Pi=\frac{15}{8}\simeq1.875
\qquad (T>\TR,\; \text{perturbative reheating}) .
\label{eq:Pi_15_8}
\end{equation}
After reheating completes, $S$ is conserved and $\Pi=0$.
We define $X>0$ to denote a positive charge excess; for $X=B$ or $B-L$, this is the convention corresponding to matter over antimatter. We adopt the convention that $\mu_X>0$ biases equilibrium toward positive $X$.
Since $\KernX(t)\ge0$ and $\Pi(t)\ge0$ during the entropy-producing interval, the entropy-clock contribution obeys
$\operatorname{sgn}Y_X=\operatorname{sgn}\epsilon_X$ within this convention; reversing $\epsilon_X$ reverses the asymmetry.
The normalization conditions below therefore constrain $|\epsilon_X\Pieff|$, not its sign.
This immediately creates the main consistency condition: entropy production must overlap the epoch in which the $X$-violating interaction is freezing out.

Combining Eqs.~\eqref{eq:master_kernel} and \eqref{eq:entropy_clock},
\begin{equation}
Y_X(\infty)=\kappa_X\epsilon_X\int dt\,\KernX(t)\Pi(t)\frac{H(t)}{T(t)} .
\label{eq:overlap}
\end{equation}
If $H/T$ varies mildly over the effective kernel, define
\begin{equation}
\Pieff\equiv\int dt\,\KernX(t)\Pi(t)\equiv\zeta\,\Pi_{\rm RH},
\qquad 0\le\zeta\le1,
\label{eq:Pieff_zeta}
\end{equation}
where $\Pi_{\rm RH}=15/8$ for the simple perturbative reheating benchmark and $\zeta$ is an overlap efficiency.
Then
\begin{equation}
Y_X(\infty)\simeq\kappa_X\epsilon_X\Pieff\left.\frac{H}{T}\right|_{\Tov} .
\label{eq:slow_overlap}
\end{equation}
For radiation domination,
\begin{equation}
\left.\frac{H}{T}\right|_{\rm RD}=1.66\sqrt{g_*}\frac{T}{M_{\rm Pl}}
=1.36\times10^{-6}\left(\frac{g_*}{100}\right)^{1/2}
\left(\frac{T}{10^{12}\,\mathrm{GeV}}\right).
\label{eq:HT_RD}
\end{equation}
If the overlap occurs before reheating completes, a conservative matter-dominated estimate is
\begin{equation}
\left.\frac{H}{T}\right|_{\rm RH}
\simeq
\left.\frac{H}{T}\right|_{\rm RD}
\left(\frac{T}{\TR}\right)^2,
\qquad T>\TR,
\label{eq:HT_RH}
\end{equation}
up to order-unity factors depending on the reheating convention and on $g_*(T)$.
Thus the radiation-dominated normalization below should be interpreted as a conservative reference value when the actual overlap is slightly before completion.

\begin{figure}[t]
\centering
\includegraphics[width=0.92\columnwidth]{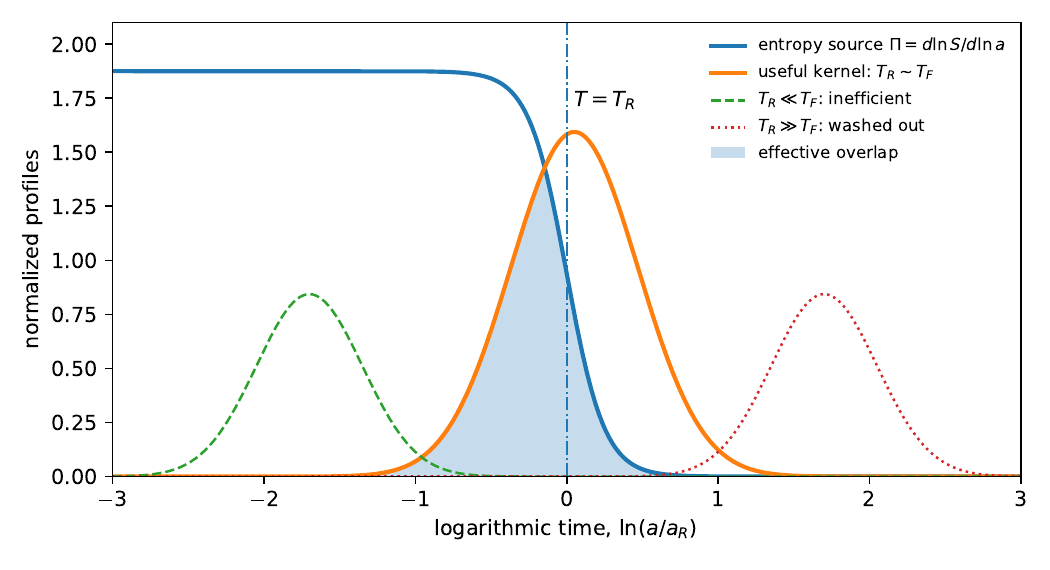}
\caption{Schematic reheating-overlap problem. Entropy production is nonzero before reheating completes and shuts off at $\TR$. A charge-violating freeze-out kernel centered near $\TF$ samples the entropy-clock source only if the two epochs overlap. If $\TR\gg\TF$, entropy production is over before freeze-out and the source is washed out. If $\TR\ll\TF$, the interaction is inefficient throughout reheating. The useful near-completion regime has $\TR={\cal O}(\TF)$; its precise width requires the full kinetic solution.}
\label{fig:overlap_SM}
\end{figure}

\section{Numerical target and Weinberg-operator benchmark}
Relating the produced charge to baryon number by $Y_B^{\rm obs}=a_XY_X$ gives, using Eq.~\eqref{eq:HT_RD},
\begin{equation}
\left|\epsilon_X\Pieff\right|\frac{\Tov}{M_{\rm Pl}}
=\frac{Y_B^{\rm obs}}{a_X\,1.66\sqrt{g_*}\,\kappa_X}
\simeq
\frac{2.6\times10^{-10}}{a_X}
\left(\frac{0.02}{\kappa_X}\right)
\left(\frac{100}{g_*}\right)^{1/2},
\label{eq:constraint}
\end{equation}
where $Y_B^{\rm obs}\simeq8.7\times10^{-11}$~\cite{Planck:2018vyg}.
For a direct baryon source, $a_X=1$.
For a $B-L$ source processed by sphalerons,
\begin{equation}
Y_B=\asph Y_{B-L},
\qquad
\asph=\frac{28}{79},
\label{eq:asph}
\end{equation}
so $a_X=\asph$~\cite{Harvey:1990qw}.
For $g_*=100$ and $\kappa_X=0.02$,
\begin{align}
\left|\epsilon_X\Pieff\right| &\simeq 3.2\times10^{-3}
\left(\frac{10^{12}\,\mathrm{GeV}}{\Tov}\right)
&& \text{direct }B,\label{eq:direct_num}\\
\left|\epsilon_X\Pieff\right| &\simeq 9.0\times10^{-3}
\left(\frac{10^{12}\,\mathrm{GeV}}{\Tov}\right)
&& B-L\text{ with sphaleron conversion}.\label{eq:bl_num}
\end{align}
With $\Pieff=\zeta(15/8)$ this becomes
\begin{align}
\left|\epsilon_X\right|\zeta &\simeq 1.7\times10^{-3}
\left(\frac{10^{12}\,\mathrm{GeV}}{\Tov}\right)
&& \text{direct }B,\label{eq:eps_zeta_B}\\
\left|\epsilon_X\right|\zeta &\simeq 4.8\times10^{-3}
\left(\frac{10^{12}\,\mathrm{GeV}}{\Tov}\right)
&& B-L\text{ with sphaleron conversion}.\label{eq:eps_zeta_BL}
\end{align}
These numbers are order-of-magnitude targets for model building, not precision predictions.
If $\Tov>\TR$ during reheating, Eq.~\eqref{eq:HT_RH} lowers the required $|\epsilon_X\Pieff|$ by the factor $(\TR/\Tov)^2$ relative to the radiation-dominated reference estimate.

A concrete benchmark is $X=B-L$ with the Weinberg operator $(LH)^2/\Lambda_\nu$.
The rate of $\Delta L=2$ scatterings can be written schematically as~\cite{Davidson:2008bu}
\begin{equation}
\Gamma_{\Delta L=2}(T)\simeq c_\nu\frac{\bar m_\nu^2T^3}{v^4},
\qquad v=246\,\mathrm{GeV},
\qquad c_\nu={\cal O}(10^{-1}).
\end{equation}
Equating this rate to the Hubble rate in radiation domination gives
\begin{equation}
\TF\sim10^{12}\text{--}10^{13}\,\mathrm{GeV}
\left(\frac{0.05\,\mathrm{eV}}{\bar m_\nu}\right)^2.
\label{eq:TF}
\end{equation}
During perturbative reheating, using Eq.~\eqref{eq:HT_RH},
\begin{equation}
\frac{\Gamma_{\Delta L=2}}{H_{\rm RH}}(T)
\simeq
\left(\frac{T}{\TF}\right)
\left(\frac{\TR}{T}\right)^2
=\frac{\TR^2}{\TF T},
\qquad T>\TR .
\label{eq:ratio_RH}
\end{equation}
The maximum of this ratio during reheating occurs near $T\simeq\TR$ and is approximately $\TR/\TF$.
Therefore, if $\TR\ll\TF$, the Weinberg interaction is never efficient during entropy production.
Conversely, if $\TR\gg\TF$, entropy production has ended long before the radiation-dominated freeze-out epoch; charge generated earlier is exposed to subsequent washout while $\Gamma_{\Delta L=2}\gg H$.
The viable entropy-clock window is therefore the near-completion overlap regime
\begin{equation}
\TR={\cal O}\!\left[\TF(\bar m_\nu)\right],
\label{eq:TR_TF_condition}
\end{equation}
with the exact width determined by the full kinetic solution.
This is a prediction/consistency condition, not a generic feature of all reheating histories.
It links the neutrino-mass scale to the reheating scale and, indirectly, to any observable that constrains reheating.

\begin{figure}[t]
\centering
\includegraphics[width=0.94\columnwidth]{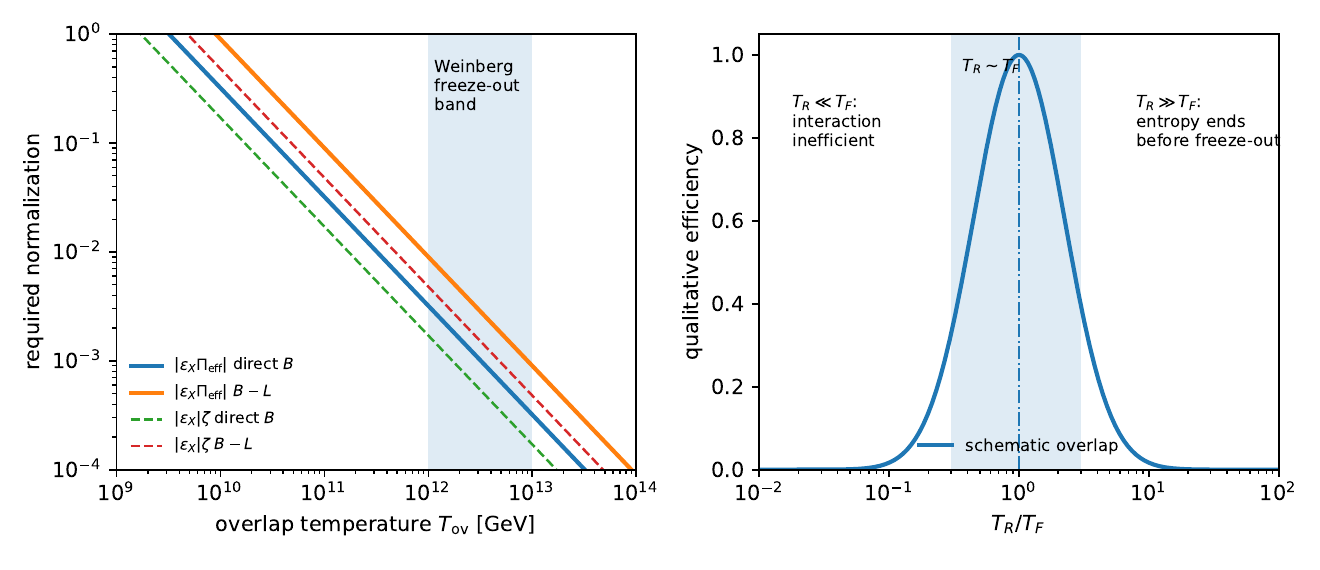}
\caption{Numerical normalization and consistency map. Left: required $|\epsilon_X\Pieff|$ in the radiation-dominated reference normalization for a direct $B$ source and for a $B-L$ source reprocessed by sphalerons. Dividing by $15/8$ gives the required $|\epsilon_X|\zeta$ in the perturbative-reheating benchmark. The shaded band indicates the typical Weinberg-operator freeze-out range for $\bar m_\nu\simeq0.05\,\mathrm{eV}$. Right: qualitative overlap condition as a function of $\TR/\TF$. The entropy-clock source is useful only in a near-completion overlap regime; far to the left the interaction is inefficient during reheating, and far to the right entropy production ends before freeze-out. The drawn efficiency profile is schematic and does not determine the regime's width.}
\label{fig:numerical}
\end{figure}

\section{Model-building status, limitations, and comparison}
A heavy scalar $\phi$ with derivative coupling $(\partial_\mu\phi/\Lambda_*)J_X^\mu$ gives $\mu_X=\dot\phi/\Lambda_*$.
A coupling to the stress-energy trace,
\begin{equation}
{\cal L}\supset \frac{\phi}{f}T^\mu_{\ \mu},
\end{equation}
leads in the tracking regime $m\gg H$ to
\begin{equation}
\phi_{\rm eq}\simeq \frac{\langle T^\mu_{\ \mu}\rangle}{m^2f},
\qquad
\mu_X\simeq \frac{1}{\Lambda_*m^2f}\frac{d}{dt}\langle T^\mu_{\ \mu}\rangle .
\label{eq:trace_source}
\end{equation}
Equation~\eqref{eq:trace_source} is not the same as Eq.~\eqref{eq:entropy_clock}: $d\langle T^\mu_{\ \mu}\rangle/dt$ need not track $d\ln S/dt$ or have a fixed sign in all reheating histories.
The trace-coupled scalar is therefore only a model-building illustration of how nonadiabatic background evolution can source a chemical potential.
A related gravitational extension previously proposed a dilaton/conformal-anomaly realization~\cite{Mandel:2026unified}.
The present analysis makes explicit that trace tracking by itself does not generically derive $\dot{\ln S}$; an actual entropy-clock completion must demonstrate this dynamical relation rather than assume it.
A true entropy-clock completion must realize, over the relevant interval,
\begin{equation}
\frac{\phi_{\rm eff}(t)}{\Lambda_*}\simeq \epsilon_X\ln\!\left[\frac{S(t)}{S_0}\right]
\end{equation}
or an equivalent monotonic dissipative variable.

Table~\ref{tab:comparison} summarizes the distinction from nearby mechanisms.
The entropy-clock form can work at $R=0$ because it is not a curvature source, but this advantage is paid for by the need to explain why a charge-biasing phase with suitable CP properties follows entropy production.
It also has a sharper timing requirement: $\GammaS\neq0$ and the charge-violating kernel must overlap.

\begin{table}[t]
\caption{Comparison of related spontaneous-baryogenesis mechanisms.}
\label{tab:comparison}
\begin{ruledtabular}
\begin{tabular}{lccc}
Mechanism & Source & Main advantage & Main limitation \\
\hline
Cohen--Kaplan & $\dot\phi/\Lambda_*$ & standard spontaneous-baryogenesis form & scalar dynamics, backreaction, isocurvature \\
Gravitational & $\dot R/M_*^2$ & geometric source & vanishes in exact RD; higher-derivative stability issues \\
Entropy clock & $\epsilon_Xd\ln S/dt$ & sign-definite during entropy production & phenomenological source; requires $\TR={\cal O}(\TF)$ overlap \\
\end{tabular}
\end{ruledtabular}
\end{table}

The limitations are therefore explicit.
First, the entropy-clock source is not derived from a unique UV completion.
Second, in a Weinberg-operator benchmark the mechanism is viable only if the reheating scale is near the $B-L$ freeze-out scale.
Third, late entropy injection is bounded by Big Bang Nucleosynthesis, requiring reheating to complete above the MeV scale~\cite{Cyburt:2015mya}.
Fourth, high-scale reheating is bounded by the inflationary energy scale.
These restrictions are not cosmetic; they make the mechanism testable and easy to falsify in explicit reheating models.

\section{Conclusions}
We derived a transfer-function criterion for baryogenesis from time-dependent derivative sources that generate an effective charge bias.
A zero-mean oscillatory source convolved with a smooth finite-time production-survival kernel is suppressed when the source varies faster than the kernel.
The exponential shut-off result $F_{\rm amp}=1/\sqrt{1+\omega^2\tau_{\rm off}^2}$ is a solvable envelope, while the integration-by-parts inequality gives the general reason for the high-frequency suppression.

We then corrected the finite-time kernel to include entropy dilution and studied the entropy-clock ansatz $\mu_X=\epsilon_Xd\ln S/dt$.
For perturbative reheating $\Pi=15/8$, so $\Pieff$ is not an arbitrary enhancement but an overlap-reduced ${\cal O}(1)$ quantity.
The central result is the timing condition.
For a Weinberg-operator $B-L$ benchmark, entropy-clock baryogenesis requires the end of reheating to occur near the $B-L$ freeze-out scale, $\TR={\cal O}[\TF(\bar m_\nu)]$.
If $\TR\gg\TF$, entropy production is over before freeze-out; if $\TR\ll\TF$, the interaction never becomes efficient during reheating.
The mechanism therefore requires an order-unity reheating/freeze-out overlap in this benchmark rather than predicting a generic entropy-production effect; the exact width remains to be obtained from a full kinetic solution.

With this limitation stated, the useful output is a compact diagnostic: successful baryogenesis from a finite source requires a nonzero low-frequency component across the corrected production-survival kernel, and entropy-clock baryogenesis additionally requires overlap between entropy production and active charge violation.
A UV completion that makes a charge-biasing phase with appropriate CP properties track $\ln S$ remains the main open model-building requirement.

\appendix

\section{Derivation of the diluted kernel}
Start from
\begin{equation}
\dot n_X+3Hn_X=-\Gamma_X\bigl(n_X-n_X^{\rm eq}\bigr),
\qquad
n_X^{\rm eq}=s\kappa_X\frac{\mu_X}{T}.
\end{equation}
For $Y_X=n_X/s$ and $S=a^3s$,
\begin{equation}
\dot s+3Hs=s\frac{d\ln S}{dt}=s\GammaS .
\end{equation}
Therefore
\begin{equation}
\dot Y_X
=\frac{\dot n_X+3Hn_X}{s}-Y_X\frac{\dot s+3Hs}{s}
=-\Gamma_X\left(Y_X-\kappa_X\frac{\mu_X}{T}\right)-\GammaS Y_X,
\end{equation}
which gives Eq.~\eqref{eq:yield_with_entropy}.
The formal solution with $Y_X(-\infty)=0$ is Eq.~\eqref{eq:master_kernel}.

\section{Phase-resolved transfer function}
To isolate the transfer effect, introduce the normalized effective kernel-envelope product
\begin{equation}
\widehat f_{\tau}(t)\equiv
\frac{\KernX(t)A(t)}{\int_0^\infty dt'\,\KernX(t')A(t')}
=\frac{1}{\tau_{\rm off}}e^{-t/\tau_{\rm off}}\Theta(t).
\end{equation}
Thus $\tau_{\rm off}$ is the effective width of the combined production-survival kernel and source envelope, rather than a separate assumption about either factor alone. The normalized response is
\begin{equation}
I_\varphi(x)=\int_0^\infty dt\,\widehat f_{\tau}(t)\cos(\omega t+\varphi).
\end{equation}
Using
\begin{equation}
\int_0^\infty dt\,\widehat f_{\tau}(t)e^{i\omega t}
=\frac{1}{1-i\omega\tau_{\rm off}},
\end{equation}
one obtains
\begin{equation}
I_\varphi(x)=\mathrm{Re}\left[\frac{e^{i\varphi}}{1-ix}\right]
=\frac{\cos\varphi-x\sin\varphi}{1+x^2}.
\end{equation}
Maximizing over $\varphi$ gives Eq.~\eqref{eq:Fx}.

\section{Perturbative reheating scalings}
For a coherently oscillating inflaton with equation of state $w\simeq0$, the early perturbative reheating solution has $\rho_\phi\propto a^{-3}$ and a subdominant radiation bath with $T\propto a^{-3/8}$ before completion~\cite{Allahverdi:2010xz}.
Then
\begin{equation}
S=a^3s\propto a^3T^3\propto a^{15/8},
\qquad
\Pi=\frac{d\ln S}{d\ln a}=\frac{15}{8}.
\end{equation}
The Hubble rate as a function of temperature scales as
\begin{equation}
H_{\rm RH}(T)\simeq H_{\rm RD}(\TR)\left(\frac{T}{\TR}\right)^4,
\end{equation}
so
\begin{equation}
\frac{H_{\rm RH}(T)}{T}\simeq
1.66\sqrt{g_*}\frac{T}{M_{\rm Pl}}
\left(\frac{T}{\TR}\right)^2 .
\end{equation}
These relations justify Eqs.~\eqref{eq:Pi_15_8} and \eqref{eq:HT_RH} up to order-unity convention-dependent factors.

\section{General kernel bound}
Let
\begin{equation}
Y_X={\cal A}_0\int dt\,f(t)e^{i\omega t},
\qquad
f(t)=\KernX(t)A(t),
\end{equation}
with $f(t)$ smooth, integrable, and vanishing at the endpoints.
Then
\begin{equation}
\int dt\,f(t)e^{i\omega t}
= -\frac{1}{i\omega}\int dt\,f'(t)e^{i\omega t},
\end{equation}
so
\begin{equation}
|Y_X|\le\frac{|{\cal A}_0|}{\omega}\int dt\,|f'(t)|.
\end{equation}
For a kernel of width $\tau_K$, this is parametrically $|Y_X|\lesssim |{\cal A}_0|/(\omega\tau_K)$ times a kernel-dependent normalization.
Endpoint discontinuities generally generate an ${\cal O}(\omega^{-1})$ tail, whereas smoother kernels may fall faster.

\section{Gravitational-wave connection}
The condition $\TR={\cal O}[\TF(\bar m_\nu)]$ can in principle be tested indirectly.
Early matter domination, preheating, and reheating transitions can source or modify a stochastic gravitational-wave background through anisotropic stress, parametric resonance, or second-order scalar perturbations.
Conversely, a bound on the reheating temperature and duration restricts the allowed entropy-clock overlap.
This is not a unique prediction of the present ansatz, but it is a consistency channel: the same reheating history controls both $\Pieff$ and any high-frequency gravitational-wave signal associated with the end of reheating.

\end{document}